# Model-free reconstruction of magnetic correlations in frustrated magnets

Authors

**Nikolaj Roth[a], Andrew F. May[b], Feng Ye[c], Bryan C. Chakoumakos[c]* and Bo Brummerstedt Iversen[a]***

[a]Center for Materials Crystallography (CMC), Department of Chemistry and Interdisciplinary Nanoscience Center (iNANO), Aarhus University, Langelandsgade 140, Aarhus, 8000, Denmark
[b]Materials Science and Technology Division, Oak Ridge National Laboratory, Oak Ridge, TN, 37831, USA
[c]Neutron Scattering Division, Oak Ridge National Laboratory, Oak Ridge, TN, 37831, USA

Correspondence email: chakoumakobc@ornl.gov; bo@chem.au.dk

**Synopsis** A new method for direct quantification of magnetic correlations in frustrated magnets is developed based on analysis of neutron total scattering data.

**Abstract** Frustrated magnetic systems exhibit extraordinary physical properties but quantification of their magnetic correlations poses a serious challenge to experiment and theory. Current insight into frustrated magnetic correlations relies on modelling techniques such as reverse Monte Carlo methods, which require knowledge about the exact ordered atomic structure. Here we present a method for direct reconstruction of magnetic correlations in frustrated magnets by three-dimensional difference pair distribution function analysis of neutron total scattering data. The methodology is applied to the disordered frustrated magnet bixbyite, $(Mn_{1-x}Fe_x)_2O_3$, which reveals nearest-neighbor antiferromagnetic correlations for the metal sites up to a range of approximately 15 Å. Importantly, this technique allows for magnetic correlations to be determined directly from the experimental data without any assumption about the atomic structure.

## 1. Introduction

A perfect crystal is a 3D object with complete long range atomic order. Crystals containing magnetic atoms give rise to macroscopic magnetic properties, and indeed magnetic materials are essential to the function of modern society, being used extensively for information storage, electricity generation, and in motors. Most of these materials have both long-range magnetic and atomic ordering, and their magnetic structures are quite well understood. However, advanced technologies will require more complex and even exotic magnetic phenomena, where atomically-ordered materials do not possess long range magnetic ordering. These disordered or frustrated magnetic materials include spin-glasses (Lee *et al.*, 1996, Lee *et al.*, 2002, Paddison, Ong, *et al.*, 2016), spin-liquids (Banerjee *et al.*, 2016, Banerjee *et al.*, 2017), spin ice (Fennell *et al.*, 2009, Morris *et al.*, 2009), superconductors (Glasbrenner *et al.*, 2015, Tranquada *et al.*, 1996), and multiferroics (Jang *et al.*, 2017, Kalinin, 2017, Zhou *et al.*, 2007). Such materials only contain local short-range correlations in their magnetic structures, and this makes it impossible to apply conventional experimental methods such as neutron diffraction, which is commonly used for studying long-range magnetism. Consequently, the progress on understanding and designing disordered spin systems has been hindered by the lack of adequate characterization of the local magnetic structure.

Magnetic disorder gives rise to fascinating phenomena, but in fact many crystals do not even contain 3D atomic order. Atomic disorder in itself leads to a range of exciting properties, and as an example atomic disorder strongly disrupts heat conduction in crystals. This has been used in numerous applications, such as the design of high performance thermoelectrics (Tan *et al.*, 2016).

Studies of atomic disorder represent a frontier of structural science, and the recent introduction of the three-dimensional difference pair distribution function, 3D-ΔPDF, obtained from X-ray scattering on single crystals has been a huge advance (Weber & Simonov, 2012). The 3D-ΔPDF method gives a three-dimensional view of only the disorder by eliminating contributions from the average ordered structure, which can be determined from conventional crystallographic methods. Unlike the much used one-dimensional PDF technique (Billinge & Egami, 2003), the 3D-ΔPDF method separates interactions at equal distances but different spatial directions, and it also makes observation of weak disorder possible in systems with a superimposed average order. The 3D-ΔPDF thus provides information that cannot be obtained from other experimental techniques.

Similar to the 3D-ΔPDF based on X-ray scattering, diffuse magnetic neutron scattering can be used to gain insight into spin-spin correlations in magnetically disordered materials. Traditionally, this method has mainly relied on inspection of the wave-vector and the temperature dependence of the scattering. Such an approach only gives a limited understanding of the disorder, as reciprocal space analysis makes the interpretation of results challenging in terms of real a space physical model. Recently, more advanced methods have been developed, where modelling of the scattering pattern is done by reverse Monte-Carlo simulations for both powder and single-crystal data (Paddison *et al.*, 2013, Paddison, Gutmann, *et al.*, 2016). In these methods a model crystal is built and its structure refined to obtain a good match between the calculated scattering pattern and the experimental data. Another recent approach has been application of magnetic pair distribution function (mPDF) analysis for powder neutron scattering (Frandsen *et al.*, 2014). Such analysis gives a one-dimensional representation of the pairwise magnetic interactions - both ordered and disordered. There are, however, at least two shortcomings of this 1D technique. One is for systems with an average magnetic order, but where there are local deviations. For such systems the average order will dominate the mPDF and the disorder will be difficult to observe. Another case is systems where different pairwise interactions have similar distances leading to peak overlap in 1D data. In such cases it will be highly challenging to uniquely establish the magnetic structure. Here we derive an expression for a three-dimensional magnetic difference pair distribution function (3D-mΔPDF). This function provides a model independent 3D reconstruction of magnetic disorder in real space. Since it does not rely on a priori information about the atomic structure it allows studies of magnetism in both atomically and magnetically disordered materials, and indeed the combination of these may lead to discovery of extraordinary new physical phenomena.

## 2. The Three-dimensional magnetic difference pair distribution function

The 3D-ΔPDF used for X-ray scattering is defined as the inverse Fourier transform of the scattered diffuse intensity, which is equal to the autocorrelation of the difference between the total electron density and the average periodic electron density, $\delta\rho(\boldsymbol{r}) = \rho_{total}(\boldsymbol{r}) - \rho_{periodic}(\boldsymbol{r})$, (Weber & Simonov, 2012):

$$\text{3D-}\Delta\text{PDF} = \mathcal{F}^{-1}\left[I_{diffuse}\right] = \langle \delta\rho \otimes \delta\rho \rangle \qquad [1]$$

where $\langle \dots \rangle$ is the experiment time-average and $\otimes$ the cross-correlation operator. The X-ray scattering 3D-ΔPDF thus only contains information about the atomic disorder, making it a powerful tool for establishing the local structure of disordered materials. The autocorrelation of the difference density will have positive peaks for vectors separating more electron density than in the average periodic structure, and negative peaks for vectors separating less electron density than the average periodic structure.

Similar to the X-ray scattering 3D-ΔPDF we define a three-dimensional magnetic difference pair distribution function as the inverse Fourier transform of the unpolarized magnetic diffuse neutron scattering cross-section.

$$\text{3D-m}\Delta\text{PDF} = \mathcal{F}^{-1}\left[\frac{d\sigma_{Diffuse}}{d\Omega}\right] \qquad [2]$$

As the interaction potential for magnetic neutron scattering is a vector field, and not a scalar field as for x-ray scattering, it is no longer simply the autocorrelation of a scalar density. We start our derivation by partitioning the magnetization density into an average periodic contribution and the deviations from it

$$\boldsymbol{M}(\boldsymbol{r}) = \boldsymbol{M}_{periodic}(\boldsymbol{r}) + \delta\boldsymbol{M}(\boldsymbol{r}) \qquad [3]$$

Note that in the case where there is no periodic magnetization density, $\boldsymbol{M}(\boldsymbol{r}) = \delta\boldsymbol{M}(\boldsymbol{r})$. We wish to express the 3D-mΔPDF in terms of this difference magnetization density. In the supporting information we show, starting from standard equations (Lovesey, 1984), that the 3D-mΔPDF can be written as:

$$\text{3D-m}\Delta\text{PDF} = \frac{r_0^2}{4\mu_B^2}\langle \delta\boldsymbol{M} \,\overline{\otimes}\, \delta\boldsymbol{M} - \frac{1}{\pi^4}(\,\delta\boldsymbol{M} \,\overline{*}\, \boldsymbol{\Upsilon}) \otimes (\delta\boldsymbol{M} \,\overline{*}\, \boldsymbol{\Upsilon})\rangle \qquad [4]$$

where we have defined the vector field cross correlation operator as a combination of element wise cross correlation and a dot product:

$$\boldsymbol{f} \,\overline{\otimes}\, \boldsymbol{g} \overset{\text{def}}{=} f_1 \otimes g_1 + f_2 \otimes g_2 + f_3 \otimes g_3 \qquad [5]$$

Where $f_i$ and $g_i$ are the vector components of $\boldsymbol{f}$ and $\boldsymbol{g}$. Similarly, we have defined the vector field convolution operator $\overline{*}$ from the scalar field convolution, $*$. The smearing function modifying the magnetization density in the second term is given by:

$$\boldsymbol{\Upsilon}(\boldsymbol{r}) = \begin{cases} \frac{\boldsymbol{r}}{|r|^4}, & |r| \neq 0 \\ \boldsymbol{0}, & |r| = 0 \end{cases} \qquad [6]$$

The first term in equation [4] is the vector autocorrelation of the difference magnetization density. Positive peaks in this function occur when the vector $\boldsymbol{r}$ separates more magnetization density pointing in the same direction than in the average periodic structure. Likewise, a negative peak occurs for vectors separating less magnetization density pointing in the same direction than in the average periodic structure. This can occur either if the magnetization direction is the same as the average, but less density is separated by the vector locally, or if the density separated by $\boldsymbol{r}$ is oppositely aligned compared with the average structure. An important simplification occurs when there is no periodic magnetic structure such as in frustrated magnets. In this case a positive peak in the first term means that the magnetization density separated by $\boldsymbol{r}$ tends to be along the same direction, and a negative peak means the magnetization density separated by the vector is oppositely aligned.

The second term in equation [4] is less straight-forward. The term originates from the fact that the scattering experiment only sees the magnetization density perpendicular to the scattering vector. A corresponding term was found by Frandsen *et al.* for the one-dimensional magnetic PDF (Frandsen *et al.*, 2014). In this term, the magnetization density is vector convoluted with the smearing function $\boldsymbol{\Upsilon}(\boldsymbol{r})$ before the autocorrelation is taken. To get a better understanding of the effect of this second term, the 3D-mΔPDF for a number of simple systems is evaluated.

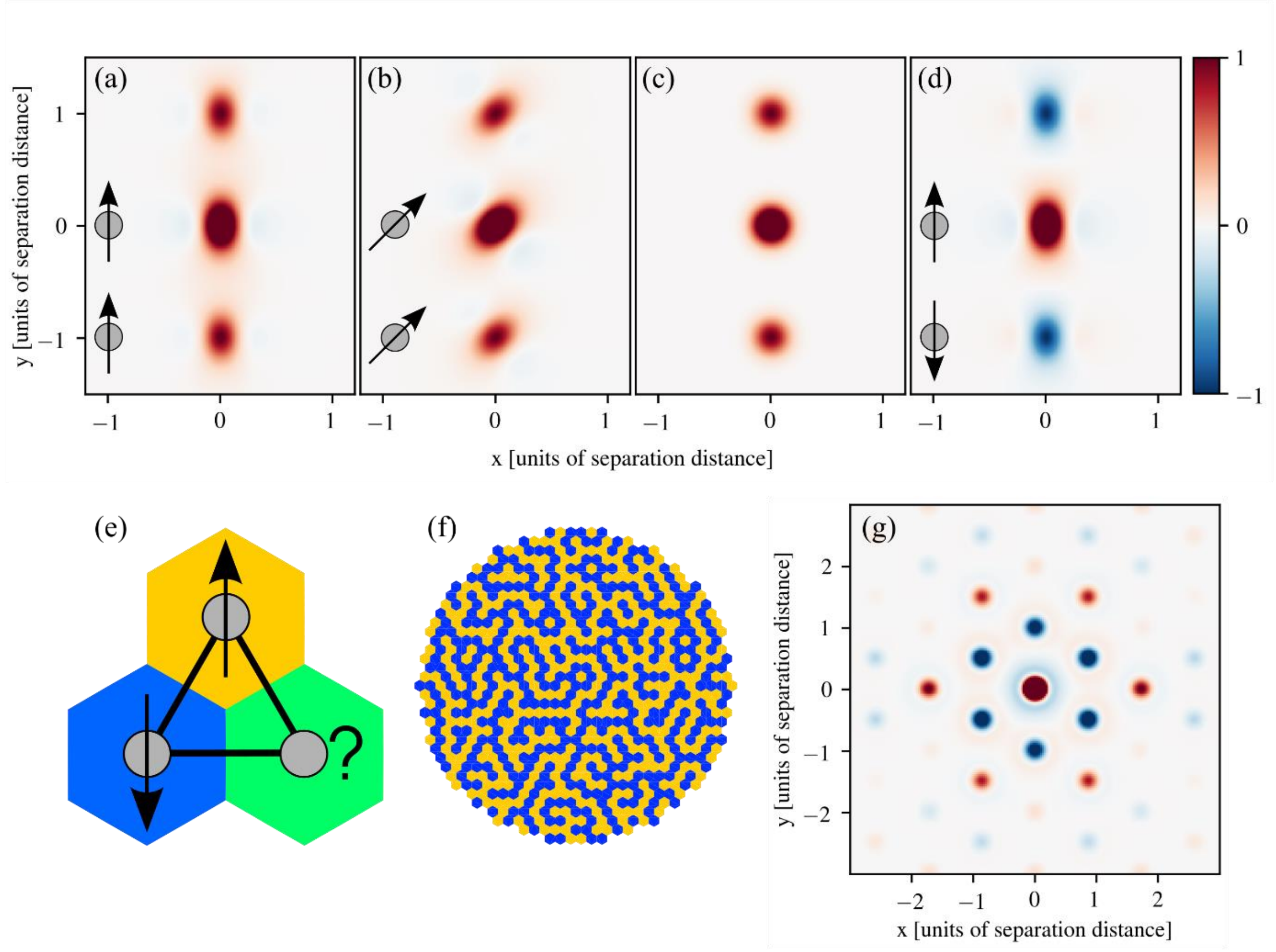

**Figure 1** Simulations of the 3D-mΔPDF for simple systems. (a) Ferromagnetic alignment along the separating axis. A positive peak is always present at the origin, as all magnetization density is aligned with itself. Positive peaks are also found at the separation vector, showing moments are aligned in same direction. (b) Ferromagnetic alignment tilted with respect to the separating axis. The 3D-mΔPDF is smeared in the direction of the moments. (c) Ferromagnetic alignment symmetry averaged for cubic symmetry. (d) Antiferromagnetically aligned moments. Negative peaks are found at the separation vector showing the opposite directions. (e) Antiferromagnetically coupled spins on a triangle. (f) A disordered ground state of the antiferromagnetic triangular Ising net (Wannier, 1950). Moments pointing into the plane are shown as blue and moments going out of the plane are yellow. It is calculated by starting with a random distribution of spin up/down, then repeatedly selecting a random spin and flipping it if it has more neighbors of the same type than opposite. (g) The 3D-mΔPDF for the antiferromagnetic triangular Ising net.

### 3. Simulations

We first simulate the 3D-mΔPDF for a system with two localized magnetic moments, modelled by Gaussian densities, in the cases where they are ferro- and antiferromagnetically coupled and aligned along different directions. Figure 1A, B and C show the 3D-mΔPDF for two moments aligned ferromagnetically. For these, as for all other 3D-mΔPDF maps, a positive peak is observed at the origin, as all magnetization density is aligned with itself. Additional positive peaks are found at the separation vector between the two moments. This shows that the moments are aligned in the same direction. The difference between A and B is the tilt of the moments relative to the separation axis. A smearing is observed in the direction of the moments, coming from the second term in equation [4]. In Figure. 1C the cubic symmetry average is shown for $m\bar{3}$ symmetry and here the direction dependent features of the moments are no longer seen, as positive and negative smearing features cancel. In Figure 1D the 3D-mΔPDF for two antiferromagnetically aligned moments is shown. Here negative peaks are found at the separation vector, showing the opposite alignment of moments.

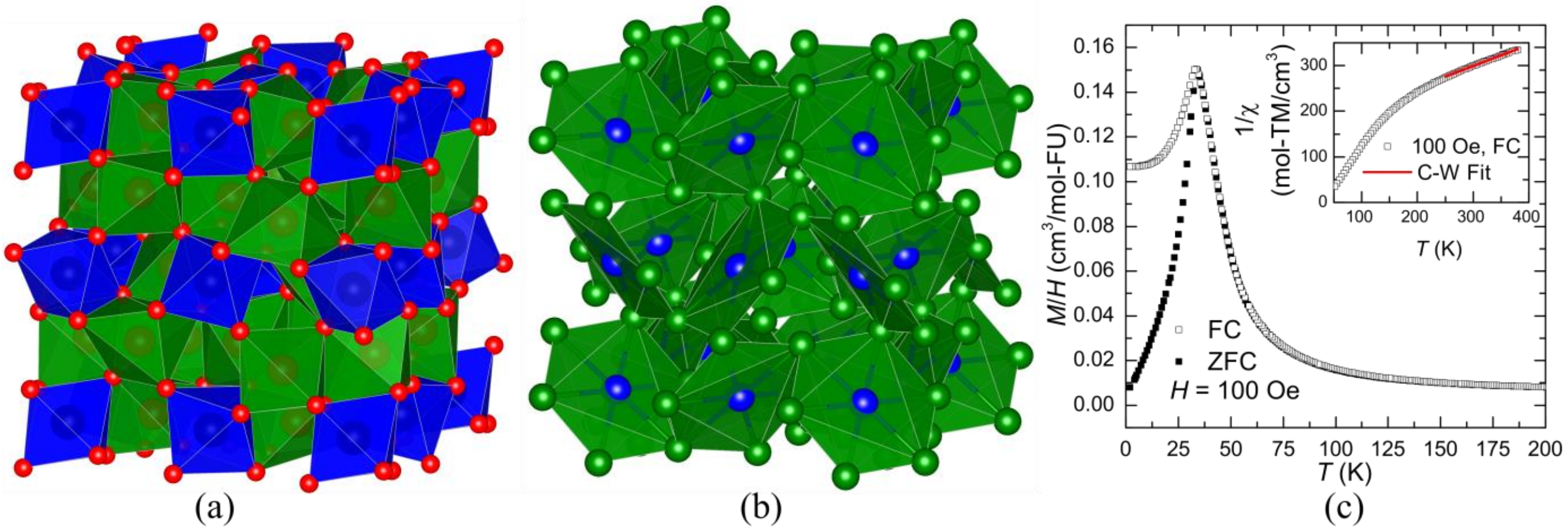

**Figure 2** Structure and magnetization of bixbyite. (a) Polyhedral model of bixbyite, where the M1 octahedra are shaded blue and the M2 polyhedra are green. The red spheres are oxygen atoms each tetrahedrally coordinated by M. The octahedra share corners and edges to make a 3D framework. (b) M-only atoms of bixbyite showing the near-neighbors of the M1 sites (blue) surrounded by the M2 sites (green). Nearly perfect hexagons of $M1(M2)_6$ result and share corners to make a 3D cubic network. (c) Field Cooled (FC) and Zero-Field Cooled (ZFC) magnetization data for bixbyite. The inset shows the $1/\chi$ behavior, with the red line indicating the Curie-Weiss fit.

The 3D-mΔPDF method is expected to be especially useful for systems with frustrated magnetism, which occurs when the magnetic moments in a structure are prohibited from having all preferences for correlations fulfilled. A simple example of this is three moments in a triangle with antiferromagnetic coupling, as shown in Fig. 1E. For such a system, it is only possible to satisfy 2 out of 3 interaction preferences. Similarly, the antiferromagnetic triangular Ising net will adapt a disordered ground state, as it is not possible for all moments to be neighboring moments of opposite direction as was shown by Wannier (Wannier, 1950). An example of one such ground state is shown in Fig 1F. In Fig. 1G we show the corresponding 3D-mΔPDF. The vectors for the nearest neighbor interactions show negative peaks, indicating the preference for antiferromagnetic alignment. Similarly, the next-nearest neighbor vectors show positive peaks, indicating that these spins to tend to align in the same direction. From the features of the 3D-mΔPDF, information about the relative orientation of magnetic moments can thus be observed directly. The interpretation of peaks is the same as for the first term in equation [4], while keeping in mind that features are smeared out due to the second term.

### 4. Experimental determination of the 3D- mΔPDF

To demonstrate the strength of our new method, we study the magnetic disorder in naturally occurring mineral bixbyite, $(Mn^{+3},Fe^{+3})_2O_3$, which has the β -$Mn_2O_3$ crystal structure (cubic, $Ia\bar{3}$, a = 9.41 Å) (Pauling & Shappell, 1930). This crystal structure has triangular and hexagonal arrangements of near-neighbor metal sites, M1 and M2, as seen in Fig. 2B, and this suggests the possibility of magnetic frustration. The naturally occurring crystal used for this study is of composition $Fe_{1.1}Mn_{0.9}O_3$ as found by both neutron diffraction and ICP measurements (See supporting information). The Fe and Mn atoms are disordered over the two metal sites in the structure. From magnetization measurements, it is found that a transition occurs at T*=32.5K, as seen as in the cusp at in Fig 2C, where temperature dependent magnetization data are shown for Field Cooled (FC) and Zero-Field Cooled (ZFC) measurements. The inset in the figure shows $1/\chi$ plotted as a function of temperature, and the red line shows the region where a Curie Weiss law is obeyed. The data clearly reveal a negative Weiss temperature, indicating that antiferromagnetic interactions are dominant in the paramagnetic phase at high temperatures. To understand the low temperature magnetic phase, single-crystal neutron scattering data were collected. The nuclear structure is identical at all temperatures in the range 7-300K, and there is no sign of long range magnetic ordering. To verify the lack of long-range magnetic ordering we have measured time, temperature and field-dependent DC magnetization, ac magnetic susceptibility and specific heat capacity, as shown in the supporting information. These measurements support the treatment of bixbyite $Fe_{1.1}Mn_{0.9}O_3$ as a phase without long-range magnetic order, and T* is found to be associated with a spin-glass transition.

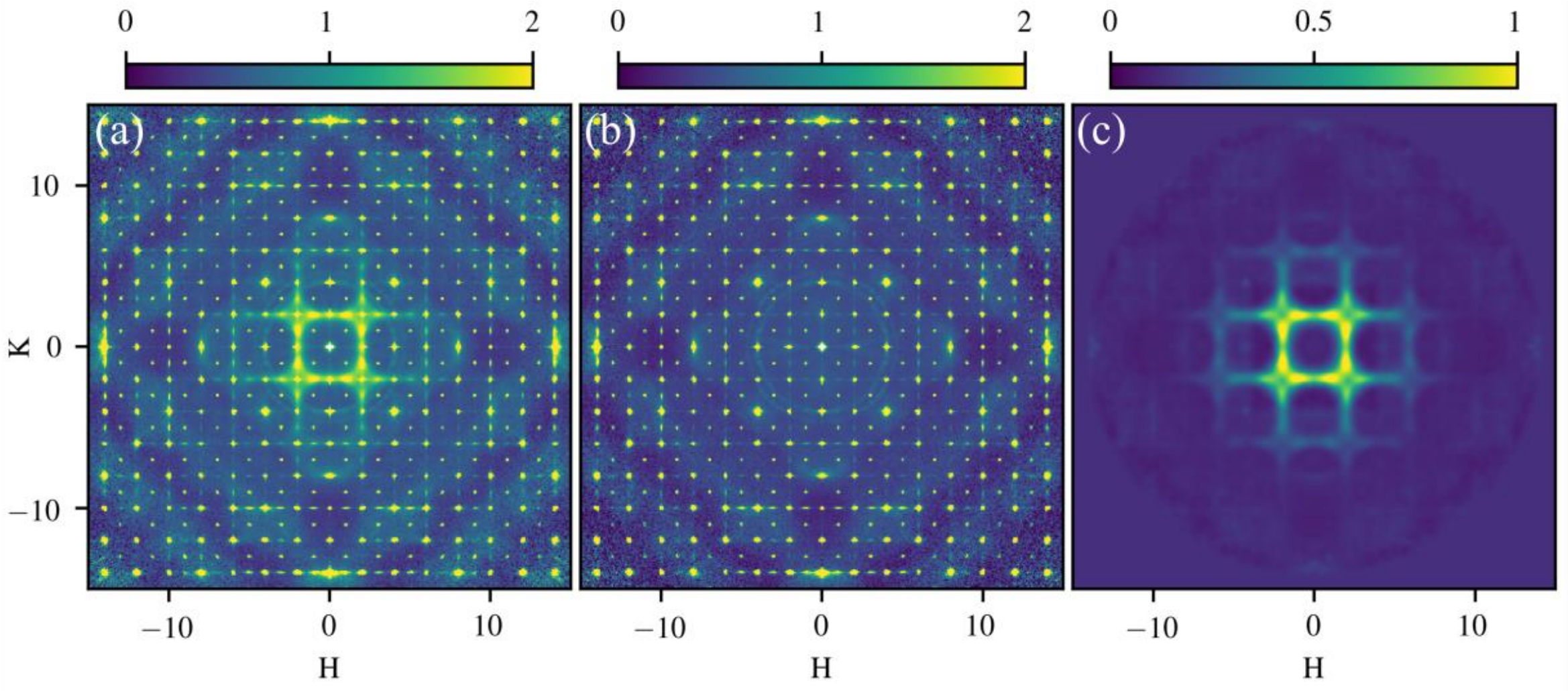

**Figure 3** Reciprocal space neutron scattering for bixbyite. All figures are of the HK0 plane. (a) Total Elastic scattering at 7K. (b) Total elastic scattering at 300K. (c) Isolated magnetic diffuse scattering.

As there is no long range magnetic order on the metal sites in the low temperature phase of bixbyite, the resulting 3D-mΔPDF will be straightforward to interpret. Because $\boldsymbol{M}_{periodic}(\boldsymbol{r}) = \boldsymbol{0}$ then $\boldsymbol{M}(\boldsymbol{r}) = \delta\boldsymbol{M}(\boldsymbol{r})$ so the resulting 3D-mΔPDF will contain information of the whole magnetization density. Furthermore, as the system has cubic symmetry, the spurious effect arising from the second term in equation [4] will cancel, as was shown by simulations in Figure 1C. Positive and negative features in the 3D-mΔPDF can then directly be interpreted in terms of magnetic moments preferring parallel and antiparallel alignment, respectively.

To produce an experimental 3D-mΔPDF, the magnetic diffuse neutron scattering has to be known. We have measured the elastic unpolarized neutron scattering at 7K and 300K, where one temperature is above the transition (i.e., the paramagnetic regime) and the other is below the transition (i.e., in the disordered spin-glass regime). These data were collected at the CORELLI spectrometer at the Spallation Neutron Source at Oak Ridge National Laboratory (Rosenkranz & Osborn, 2008). CORELLI's design enables elastical discrimination of the total scattering, i.e., the phonon and thermal diffuse scattering are removed. From the two data sets the full elastic reciprocal space scattering intensities are reconstructed using the Laue symmetry of the crystal. The reconstructed HK0 plane at the two temperatures can be seen in Fig 3A and B. Since the nuclear structure is identical at 7 K and 300 K, the data from the paramagnetic regime can be subtracted from the low temperature data to remove all scattering contributions other than magnetic scattering. This includes nuclear scattering, both Bragg and diffuse, as well as background scattering. After the subtraction, residual errors are present at the position of the very sharp Bragg peaks. To remove these, a punch-and-fill method is employed, where a small volume around each reflection is removed and filled with a smooth function to resemble the diffuse scattering in that region (See the supporting information). In cases where there would be a long range magnetic ordering, the same punch and fill method would be used to remove the magnetic Bragg scattering. The high-angle data are also removed, as they mainly consist of noise. The result of this process is the isolated magnetic diffuse scattering in 3D reciprocal space. Diffuse magnetic scattering for the HK0 plane from this process of the bixbyite data can be seen in Fig 3C. A more detailed description of the data reduction process can be found in the supporting information.

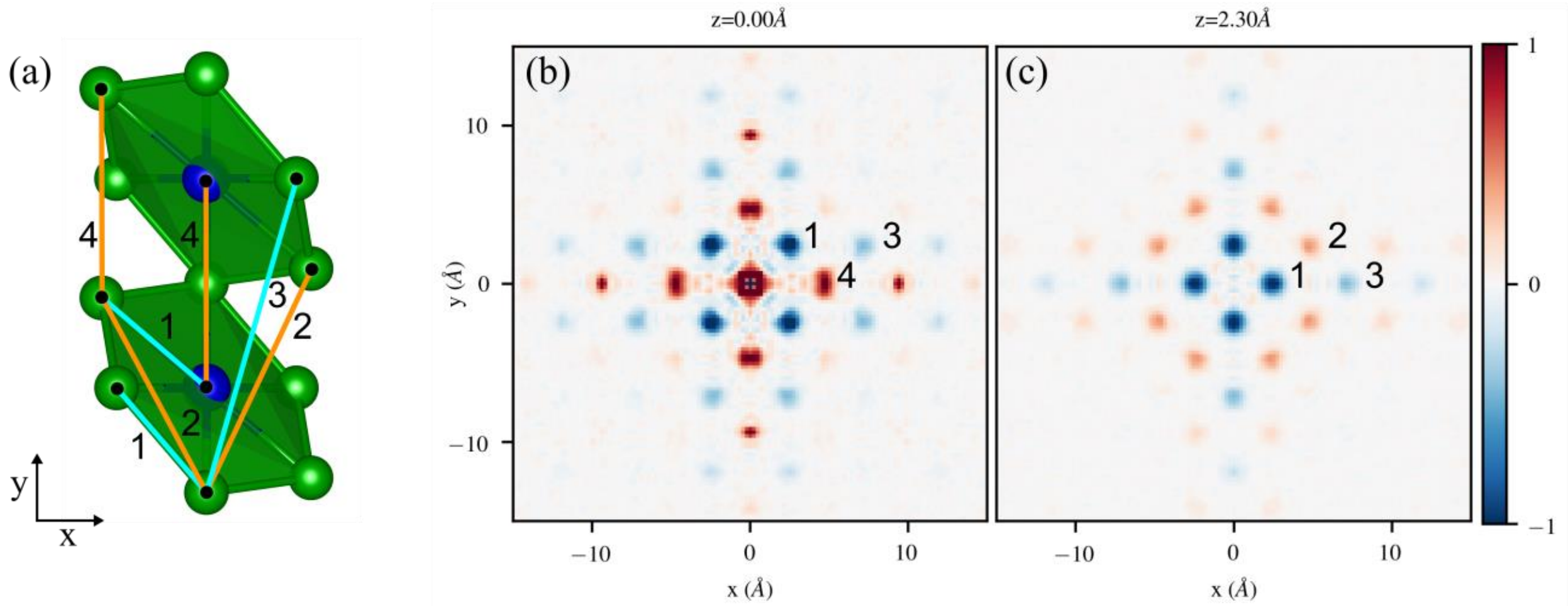

**Figure 4** The 3D-mΔPDF for bixbyite. (a) Selected portion of the structure showing numbered vectors between atoms. (b) 3D-mΔPDF for the z=0 plane. (c) 3D-mΔPDF for the z = 2.30 Å plane. The slight split of peak number 4 is an artefact, as the vector should have only a component along one axis.

The 3D-mΔPDF is then simply obtained by Fourier transformation. Two planes of the 3D-mΔPDF for bixbyite are shown in Fig 4B and C. As there is no long-range periodic magnetic order, the peaks in the 3D-mΔPDF can be directly interpreted as the alignment preference between sites separated by the corresponding vector. A few of the features in the maps have been marked with numbers for which the corresponding vectors in the crystal structure are shown in Fig. 4A. The nearest-neighbor vector (marked 1), which is both for site pairs M1-M2 and M2-M2 has a negative peak in the 3D-mΔPDF, which identifies that nearest-neighbor metal sites tend to have antiferromagnetic alignment. The 3D-mΔPDF for the vectors for the next-nearest neighbor pairs (marked 2) is positive, showing preference for alignment in the same direction as both of the metal sites in the structure contain disordered mixtures of Fe and Mn, the local magnetic structure could be expected to be very complicated, depending on local distributions of Fe and Mn on the two sites. However using the 3D-mΔPDF technique it is seen that on average the metal sites have an antiferromagnetic nearest-neighbor correlation. These correlations can then be followed to longer distances showing alternating positive and negative peaks for higher order neighbors. Overall, the 3D-mΔPDF for bixbyite clearly shows the disordered low temperature state to be dominated by an antiferromagnetic nearest-neighbor interaction. Peaks in the 3D-mΔPDF fall off rapidly and disappear after around 15 Å, directly revealing the maximum distance of the magnetic correlations, which is not greater than 2 unit cells. The 3D-mΔPDF results are expectedly self-consistent with the field and temperature dependent magnetization measurements, but moreover, directly show the 3D atomic pairwise correlations that exist in the frozen spin state, without making any assumptions about the system.

## 5. Discussion

The 3D-mΔPDF has two major advantages compared with the 1D-mPDF introduced by Frandsen *et al.* (Frandsen *et al.*, 2014). One is for systems with average long-range magnetic order and local deviations. In this case, the 3D-mΔPDF will show the deviations from average structure directly, and the average magnetic structure can then be found by separate analysis of Bragg reflections. In such systems, the 1D-mPDF superimposes the order and disordered parts of the structure, making it difficult to interpret the disorder, which is often a small deviation from the average structure. The second advantage arises from the fact that the magnetic scattering falls off rapidly in reciprocal space, as the electrons responsible for the magnetic moment are diffuse. This affects the broadness of peaks in the PDF functions when Fourier transforming. For the 1D-mPDF the peak broadness can easily lead to overlap of neighboring peaks, once again making the interpretation less straight forward. Peak overlap is also obtained when multiple interactions have the same distance, but different spatial directions. In such cases the 3D-mΔPDF retains the directional information, making it possible to separate peaks close or equal in distance, but with different directions.

In the bixbyite system we were able to separate the magnetic diffuse scattering from nuclear diffuse scattering and scattering from the sample environment by subtracting a high temperature dataset of the same structure. In cases where there is a structural transition between the paramagnetic and frustrated magnetic states, this method cannot be used. In cases where there is a structural change, but the low temperature structure is ordered (no nuclear diffuse scattering), the magnetic diffuse scattering can be isolated by subtracting the scattering from an empty sample environment and using the punch and fill method on the Bragg peaks. In cases where there is a structural transition to a disordered structure, the magnetic diffuse scattering can potentially be isolated by polarized neutron scattering. Here we used the elastic discrimination of the CORELLI instrument to remove phonon scattering from the data, as this contribution is different at 300K and 7K. The effect of only using the elastic scattering on the 3D-mΔPDF is that only the static magnetic correlations are seen. This means that magnetic excitations such as magnons are not seen. For Bixbyite this is not a problem, as the magnetic diffuse scattering is elastic (see supporting information), but in cases where the dynamical magnetic correlations are wanted, the energy-integrated signal should be used for producing the 3D-mΔPDF. This is further discussed in the supporting information.

In conclusion, we have derived an expression for the magnetic three-dimensional difference PDF, 3D-mΔPDF, which directly reveals magnetic correlations for systems with disordered magnetism. Since it is a direct space function, an intuitive interpretation is easily obtained, and this provides a better understanding of magnetic disorder, even for complex systems. Unlike previous studies of disordered magnetic systems, this new method is completely model independent. As the 3D-mΔPDF is simply the Fourier transform of the magnetic diffuse scattering, it provides a direct space view of all information about the magnetic disorder contained in the scattering data. In contrast to reverse Monte Carlo models previously used for interpretation of magnetic diffuse neutron scattering, the 3D-mΔPDF is not challenged by false minima, although this can be mitigated by repeated simulations with randomized starting conditions. More importantly the 3D-mΔPDF approach does not require a specific structural model, and this makes it possible to study magnetism also in atomically disordered systems such as bixbyite. The end members of bixbyite, $Mn_2O_3$ and β-$Fe_2O_3$ are known to go through phase transitions to ordered magnetic phases (Cockayne *et al.*, 2013, Malina *et al.*, 2015). This suggests that the presence of atomic disorder allows tuning this complex magnetic system to create the magnetic frustration described above.

**Acknowledgements** The work was support by the Danish National Research Foundation (DNRF93). The authors gratefully acknowledge Bent Ørsted and Kasper Tolborg for fruitful discussions. A portion of this research used resources at the Spallation Neutron Source, a DOE Office of Science User Facility operated by the Oak Ridge National Laboratory. Physical property characterization (A.F.M.) was supported by the U.S. Department of Energy, Office of Science, Basic Energy Sciences, Materials Sciences and Engineering Division

# Supporting information

## S1. Materials and Methods

### S1.1. Physical property measurements

Using a single crystal, DC magnetization measurements were performed in a Quantum Design Magnetic Property Measurement System (MPMS), and ac magnetic susceptibility and specific heat capacity measurements were performed in a Quantum Design Physical Property Measurement System (PPMS). The ac magnetization measurements were performed with zero applied dc field, using amplitude of $A_{AC}$ = 14Oe at various frequencies $f$. Time-dependent measurements were performed in the MPMS by ramping the magnetic field from 0 to 1000 Oe at various temperatures after the sample was taken to 50K or higher and cooled in zero applied field. The quantity $M_0$ is defined as the first measurement after the applied field of 1000 Oe was stabilized, for a given temperature; note that it took roughly 150s for the field to stabilize and the origin t=0 is defined by the first measurement $M_0$ not the time when H=0.

The physical properties results are shown in figure S1. The magnetization data of Bixbyite, shown in figure S1a, are characterized by a high-temperature paramagnetic region, and a cusp at T*=32.5K that at first glance would appear to be associated with antiferromagnetic ordering. Below T*, we observe a large divergence between the zero-field cooled (warming) data and the field-cooled data, which is expected for spin-glasses (Binder & Young, 1986). To further examine this behavior, we performed specific heat capacity measurements, shown in figure S1c, and did not observe any anomaly in the specific heat at T*. AC magnetic susceptibility measurements and time-dependent dc measurements were performed to establish the presence of glassy dynamics of the spins in Bixbyite. The maximum in the in-phase component of the ac susceptibility, $\chi'$ shown in figure S1b, clearly shifts downward in magnitude and to higher temperature as the frequency $f$ increases, which is the behavior expected for a spin glass (Binder & Young, 1986, Mydosh, 1993, Balanda, 2013). Consistent with the behavior of $\chi'$ for a spin-glass, we observe a time-dependence of the dc magnetization data at T = 2 and 30K, but not at 50K, which is shown in figure S1d. Therefore, T* appears to be associated with a spin freezing transition and these results support the treatment of Bixbyite Fe1.1Mn0.9O3 as a phase that does not possess long-range order.

In the insert in figure S1a, $1/\chi$ is plotted as a function of temperature. Above 200K, the DC susceptibility $\chi = \frac{M}{H}$ is found to follow the Curie-Weiss law $\chi = \frac{C}{T-\Theta}$, where C is the Curie constant related to the effective moment and Θ the Weiss temperature. For an applied field of H=100 Oe, we fit the data between 200 and 380K and obtained an effective moment of $\frac{4.1(1)\mu_B}{TM}$, where TM = transition metal. The data also clearly reveal a negative Weiss temperature, with the fit yielding $\Theta = -336(2)K$. The negative Weiss temperature indicates antiferromagnetic interactions are dominant in the paramagnetic phase, consistent with the spin-spin correlation information obtained from the 3D-mΔPDF at lower temperatures. The frustration ratio, $|\Theta|/T^*$ = 336/32.5=10.3 demonstrates a significant amount of frustration in this natural bixbyite sample. The DC M/H data show a strong deviation from Curie-Weiss behavior below approximately 150K. From the data, it appears that $M/H$ is being enhanced ($\frac{1}{\chi}$ is reduced below the paramagnetic expectation). This may suggest some ferromagnetic component associated with the short-range order, such as canting of the local AFM order.

**S1.2. Scattering and production of 3D-mΔPDF**

Single crystal elastic neutron scattering was measured at the CORELLI beamline at the Spallation Neutron Source at Oak Ridge National Laboratory (Rosenkranz & Osborn, 2008). A piece was cut from a large cubic crystal, and subsequently sanded down to a sphere to limit crystal shape effects in the scattering. The crystal was glued to the end of an aluminum pin which was wrapped in neutron absorbing Cd-foil and mounted in a low background closed-cycle refrigerator. Measurements were carried out at 7K, 25K, 50K, 80K, 160K, 240K and 300K, with the 7K and 300K being longer for increased counting statistics. Background scattering for subtraction was measured on an aluminum pin with Cd foil, but no crystal.

For every rotation angle of the crystal, the UB matrix was determined, as it was found that small errors were present in the goniometer rotation angle. Elastic-only scattering intensities in reciprocal space were reconstructed from the data using the UB matrices and normalized to vanadium flux. Symmetrization with the Laue symmetry (*m-3*) was employed in the reconstruction to fill out reciprocal space and to increase counting statistics. If the intensities were to be used for the production of a 3D-mΔPDF, no background subtraction was employed, as it would be subtracted later. For the production of an intensity map, background scattering was subtracted. Scattered intensities were reconstructed on a 501*501*501 point grid with each axis spanning between +-13.67Å$^{-1}$.

The production of a 3D-mΔPDF is done in the following way: Intensities measured at 300K, well into the paramagnetic regime, are subtracted from intensities measured at a lower temperature. In this way, all scattering contributions other than magnetic scattering are approximately removed. This includes nuclear scattering, both Bragg and diffuse, as well as background scattering. While the broad nuclear diffuse scattering and background scattering are well subtracted, the very sharp Bragg scattering has some residual errors. To remove these, a punch and fill method is employed. A similar method was described in (Kobas *et al.*, 2005). The method used here is slightly different from the one described by Kobas..

Instead of using rectangular boxes as the punch, the closest approximation to an ellipsoid on a rectangular grid is used. In this case where a cubic grid is used for the scattered intensities, the closest approximation to a sphere is used. As the Bragg peaks are close to being spherical in reciprocal space, this type of punch allows for complete removal of peaks while removing as little as possible from the rest of the scattered intensity. Strong reflections were punched with a sphere of 7 pixels in diameter, while weak reflections were punched with spheres of 3 pixels diameter. The intensities are punched by setting their values to NaN. The punched intensities then need to be filled in by a smooth function, resembling the diffuse scattering at the Bragg peaks. We accomplish this using the astropy convolution function for Python (Robitaille *et al.*, 2013). This function allows for convolution of 3D arrays where NaN values are replaced with interpolated values using the kernel as an interpolation function. In this case we use a Gaussian with a 2 pixel standard deviation as a kernel. By the convolution function used to fill in the punched intensities, the whole scattered intensity is also smoothed slightly, as seen in Figure 3. Regions of reciprocal space where no or very small amounts of data had been measured were removed. This includes the far reciprocal space where $\sqrt{h^2 + k^2 + l^2} > 14$, the corners where h,k,l are all larger than 4.5 and a sphere of diameter 11 pixels in the center of reciprocal space, as the direct beam is not measured. Through this process, the magnetic diffuse scattering has been isolated. The process is illustrated in Fig. S2. Figure S2A and B show the elastic neutron scattering at 7K and 300K, respectively. Figure S2C shows the difference when subtracting the 300K data from the 7K data. Figure S2D shows the difference scattering with holes at the Bragg positions as well as removed high angle noise. Figure S2E shows the finished isolated magnetic diffuse scattering after the filling process.

The punch and fill method used here will have two effects on the 3D-mΔPDF, one of which is easily removed. As the diffuse scattering intensity at the Bragg positions is filled by a smooth function, some error is introduced. However as these errors will be in the high-frequency components of the intensity, it will only affect the 3D-mΔPDF for long distances. But as we are only concerned with short-range interactions in the 3D-mΔPDF, the error introduced by this is not critical. The second effect comes from the convolution of the whole scattering pattern with a Gaussian. When the array is then Fourier transformed, the resulting array will correspond to the product of the 3D-mΔPDF with the Fourier transform of the Gaussian kernel, as required by the convolution theorem. The effect of the Gaussian kernel is then simply removed by dividing the array by the Fourier transform of the Gaussian, which is known, and thereby obtaining the 3D-mΔPDF.

When subtracting the high temperature dataset from the low temperature data to remove the nuclear and background contribution from the scattering, a paramagnetic scattering signal is also subtracted. In the paramagnetic regime, the spins are not correlated and the resulting scattering is simply proportional to the single-atom scattering factor. This is very slowly varying and isotropic in reciprocal space (Lovesey, 1984). The result of this subtraction is that slightly negative values are found in the inner region of reciprocal space, as can be seen in Figure S2C. When Fourier transforming to the 3D-mΔPDF this only affects the r=0 signal and not the correlations which we are interested in.

Bragg peaks were integrated using an instrument specific script, corrected for absorption and merged using XPREP (Sheldrick, 2001) to *m*-3 symmetry. The structure was refined using SHELXL (Sheldrick, 2008) in the space group *Ia*-3. Both metal sites were refined as a combination of Mn and Fe with the constraint that each site is fully occupied. The twin law [1, 0, 0; 0, 0, 1; 0, 1, 0] was used during refinement. In Table T1 refinement results are shown.

As the fit is good, and approximately equally good at all temperatures is seems that the nuclear structure is sufficient to model the Bragg reflections. This suggests that there is no long-range magnetic ordering at low temperatures, which agrees well with the physical property measurements. To further test this, we have followed the reflection intensities with temperature. No new reflections are observed when cooling the sample, so any long-range magnetic structure would have to be contained in the existing Bragg reflections. In Figure S3 we have plotted a number of normalized low order reflection intensities with temperature, as any magnetic structure is expected to be seen in the low order region.

No significant changes in intensities are found when cooling the sample. The scattering data thus indicate that there does not seem to be any long-range magnetic ordering in the sample based on three points. These points are: 1) No appearance of new diffraction peaks when cooling the sample. 2) No significant changes in low order reflection intensities. 3) The nuclear structure describes the data well without the need for any magnetic model.

To check the composition found from the neutron diffraction data we have measured Inductively Coupled Plasma Atomic Emission Spectroscopy (ICP-EAS). From this we get then Fe/Mn ratio to be 1.15(1):0.85(1), which is close to the values found from refinement of diffraction data.

## S2. Supplemantary Text

### S2.1. Derivation

As can be found in most texts on magnetic diffuse neutron scattering, e.g. (*20*), the unpolarized cross-section for a magnetic system in the static approximation can be written as:

$$\frac{d\sigma}{d\Omega} = r_0^2 \langle \boldsymbol{Q}_\perp(-\boldsymbol{k}) \cdot \boldsymbol{Q}_\perp(\boldsymbol{k}) \rangle \qquad [S1]$$

Where $r_0 = \frac{\gamma e^2}{m_e c^2}$, $\boldsymbol{k}$ is the scattering vector, $\langle\ \rangle$ is the time average of the experiment and $\boldsymbol{Q}_\perp$ is the Fourier transform, denoted $\mathcal{F}$, of the magnetization density perpendicular to $\boldsymbol{k}$**:**

$$\boldsymbol{Q}_\perp(\boldsymbol{k}) = -\frac{1}{2\mu_B} \int d\boldsymbol{r} \exp(\mathrm{i}\mathbf{k} \cdot \mathbf{r})\, \widetilde{\boldsymbol{k}} \times \left(\boldsymbol{M}(\boldsymbol{r}) \times \widetilde{\boldsymbol{k}}\right) = -\frac{1}{2\mu_B}\, \mathcal{F}\left[\widetilde{\boldsymbol{k}} \times \left(\boldsymbol{M}(\boldsymbol{r}) \times \widetilde{\boldsymbol{k}}\right)\right] \qquad [S2]$$

Here $\boldsymbol{r}$ is the real-space vector and $\widetilde{\boldsymbol{k}} = \boldsymbol{k}/|\boldsymbol{k}|$ is the unit vector in reciprocal space. $\boldsymbol{M}(\boldsymbol{r})$ is the total magnetization density, the sum of both spin and orbital magnetization densities. Another way to partition the magnetization density is into an average periodic structure without disorder and the deviations from it

$$\boldsymbol{M}(\boldsymbol{r}) = \boldsymbol{M}_{periodic}(\boldsymbol{r}) + \delta\boldsymbol{M}(\boldsymbol{r}) \qquad [S3]$$

The term in the cross-section arising from the periodic magnetization density will give rise to sharp Bragg peaks in reciprocal space. The scattering from the deviations from the average periodic structure we call the magnetic diffuse scattering, and is what were are interested in here. There will also be a third cross term in the scattering cross section between the average structure and deviations from it. This term will only exist on the reciprocal lattice points, but will be much smaller than the Bragg scattering from the average periodic structure, and is therefore usually neglected.

We wish to arrive at an expression for the inverse Fourier transform of the magnetic diffuse scattering cross-section. We first insert equation [S2] for the deviations from periodic structure into equation [S1] and rearrange:

$$\frac{d\sigma_{Diffuse}}{d\Omega} = \frac{r_0^2}{4\mu_B^2} \langle \mathcal{F}\left[\widetilde{\boldsymbol{k}} \times \left(\delta\boldsymbol{M}(\boldsymbol{r}) \times \widetilde{\boldsymbol{k}}\right)\right]^* \cdot \mathcal{F}\left[\widetilde{\boldsymbol{k}} \times \left(\delta\boldsymbol{M}(\boldsymbol{r}) \times \widetilde{\boldsymbol{k}}\right)\right] \rangle \qquad [S4]$$

Where $^*$ denotes the complex conjugate. As the Fourier transform is an integral with respect to $\boldsymbol{r}$, then $\widetilde{\boldsymbol{k}}$ is a constant vector in regards to the transform. In this case, it can then easily be shown that

$$\mathcal{F}\left[\widetilde{\boldsymbol{k}} \times \left(\delta\boldsymbol{M}(\boldsymbol{r}) \times \widetilde{\boldsymbol{k}}\right)\right] = \widetilde{\boldsymbol{k}} \times \left(\mathcal{F}[\delta\boldsymbol{M}(\boldsymbol{r})] \times \widetilde{\boldsymbol{k}}\right) \qquad [S5]$$

By using the identity for the triple cross product and that $\widetilde{\boldsymbol{k}} \cdot \widetilde{\boldsymbol{k}} = 1$ we get:

$$\widetilde{\boldsymbol{k}} \times \left(\mathcal{F}[\delta\boldsymbol{M}(\boldsymbol{r})] \times \widetilde{\boldsymbol{k}}\right) = \mathcal{F}[\delta\boldsymbol{M}(\boldsymbol{r})] - \widetilde{\boldsymbol{k}}\left(\mathcal{F}[\delta\boldsymbol{M}(\boldsymbol{r})] \cdot \widetilde{\boldsymbol{k}}\right) \qquad [S6]$$

Inserting this into equation [S4] and reducing gives:

$$\frac{d\sigma_{Diffuse}}{d\Omega} = \frac{r_0^2}{4\mu_B^2} \langle \mathcal{F}[\delta\boldsymbol{M}(\boldsymbol{r})]^* \cdot \mathcal{F}[\delta\boldsymbol{M}(\boldsymbol{r})] - \left(\mathcal{F}[\delta\boldsymbol{M}(\boldsymbol{r})] \cdot \widetilde{\boldsymbol{k}}\right)^* \cdot \left(\mathcal{F}[\delta\boldsymbol{M}(\boldsymbol{r})] \cdot \widetilde{\boldsymbol{k}}\right) \rangle \qquad [S7]$$

We now define the three-dimensional magnetic difference pair distribution function as the inverse Fourier transform of the magnetic diffuse scattering cross-section:

$$3D - m\Delta PDF = \mathcal{F}^{-1}\left[\frac{d\sigma_{Diffuse}}{d\Omega}\right] \qquad [S8]$$

To simplify expressions, we need to consider how to treat inverse Fourier transforms of dot products. We are looking for an analogue to the cross-correlation theorem. The cross correlation, $\otimes$, of two scalar fields, $f(\boldsymbol{k})$ and $g(\boldsymbol{k})$ is related to the convolution operator, $*$, by:

$$f(\boldsymbol{k}) \otimes g(\boldsymbol{k}) = f^*(-\boldsymbol{k}) * g(\boldsymbol{k})$$

For scalar fields, the convolution and cross-correlation theorems say:

$$\mathcal{F}[f \otimes g] = \mathcal{F}[f]^* \cdot \mathcal{F}[g] \qquad \text{and} \qquad \mathcal{F}[f * g] = \mathcal{F}[f] \cdot \mathcal{F}[g]$$

If we define two vector fields

$$\boldsymbol{f}(\boldsymbol{k}) = \begin{pmatrix} f_1(\boldsymbol{k}) \\ f_2(\boldsymbol{k}) \\ f_3(\boldsymbol{k}) \end{pmatrix} \text{ and } \boldsymbol{g}(\boldsymbol{k}) = \begin{pmatrix} g_1(\boldsymbol{k}) \\ g_2(\boldsymbol{k}) \\ g_3(\boldsymbol{k}) \end{pmatrix}$$

Then the inverse Fourier transform of the dot product of $\boldsymbol{f}^*$ and $\boldsymbol{g}$ will give

$$\mathcal{F}^{-1}[\boldsymbol{f}^* \cdot \boldsymbol{g}] = \mathcal{F}^{-1}[f_1^* g_1] + \mathcal{F}^{-1}[f_2^* g_2] + \mathcal{F}^{-1}[f_3^* g_3]$$

$$= \mathcal{F}^{-1}[f_1] \otimes \mathcal{F}^{-1}[g_1] + \mathcal{F}^{-1}[f_2] \otimes \mathcal{F}^{-1}[g_2] + \mathcal{F}^{-1}[f_3] \otimes \mathcal{F}^{-1}[g_3]$$

If we then define the vector cross correlation operator, $\overline{\otimes}$, such that

$$\boldsymbol{f} \,\overline{\otimes}\, \boldsymbol{g} \stackrel{\text{def}}{=} f_1 \otimes g_1 + f_2 \otimes g_2 + f_3 \otimes g_3$$

Then we have the cross-correlation theorem for vector fields as:

$$\mathcal{F}^{-1}[\boldsymbol{f}^* \cdot \boldsymbol{g}] = \mathcal{F}^{-1}[\boldsymbol{f}] \,\overline{\otimes}\, \mathcal{F}^{-1}[\boldsymbol{g}] \quad \Leftrightarrow \quad \mathcal{F}[\boldsymbol{f}]^* \cdot \mathcal{F}[\boldsymbol{g}] = \mathcal{F}\left[\boldsymbol{f} \,\overline{\otimes}\, \boldsymbol{g}\right] \qquad \text{[S9]}$$

Similarly, this can be done to define the convolution operator $\overline{*}$, giving a vector convolution theorem.

We are now ready to return to the problem at hand, which is the inverse Fourier transform of the scattering cross- section. There are two terms in the expression. The first is of the form:

$$\mathcal{F}^{-1}\left[\mathcal{F}[\delta \boldsymbol{M}(\boldsymbol{r})]^* \cdot \mathcal{F}[\delta \boldsymbol{M}(\boldsymbol{r})]\right] = \delta \boldsymbol{M}(\boldsymbol{r}) \,\overline{\otimes}\, \delta \boldsymbol{M}(\boldsymbol{r}) \qquad \text{[S10]}$$

This term is the vector autocorrelation of the disorder magnetization density. The second term is of the form:

$$\mathcal{F}^{-1}\left[\left(\mathcal{F}[\delta \boldsymbol{M}] \cdot \widetilde{\boldsymbol{k}}\,\right)^* \cdot \left(\mathcal{F}[\delta \boldsymbol{M}] \cdot \widetilde{\boldsymbol{k}}\,\right)\right] = \mathcal{F}^{-1}\left[\mathcal{F}[\delta \boldsymbol{M}] \cdot \widetilde{\boldsymbol{k}}\right] \otimes \mathcal{F}^{-1}\left[\mathcal{F}[\delta \boldsymbol{M}] \cdot \widetilde{\boldsymbol{k}}\right] \qquad \text{[S11]}$$

Note that the cross correlation in [11] is not the vector cross correlation. The function which this term is the autocorrelation of can then be rewritten:

$$\mathcal{F}^{-1}\left[\mathcal{F}[\delta \boldsymbol{M}] \cdot \widetilde{\boldsymbol{k}}\right] = \delta \boldsymbol{M} \,\overline{*}\, \mathcal{F}^{-1}[\widetilde{\boldsymbol{k}}] \qquad \text{[S12]}$$

To understand this term, we need the inverse Fourier transform of the unit vector $\widetilde{\boldsymbol{k}}$. Let us denote this function $\mathcal{F}_{\widetilde{\boldsymbol{k}}}^{-1}(\boldsymbol{r})$

$$\mathcal{F}_{\widetilde{\boldsymbol{k}}}^{-1}(\boldsymbol{r}) = \mathcal{F}^{-1}\left[\widetilde{\boldsymbol{k}}\right] = \frac{1}{(2\pi)^3} \int d\boldsymbol{k} \frac{\boldsymbol{k}}{|\boldsymbol{k}|} \exp(-i\boldsymbol{k} \cdot \boldsymbol{r}) \qquad \text{[S13]}$$

To solve this integral, we first look for symmetry in the integral. If we rotate the argument with operator $R$ and use $\cdot R\boldsymbol{r} = R^{-1}\boldsymbol{k} \cdot \boldsymbol{r}$ , $R^{-1}R = 1$ , $|R^{-1}\boldsymbol{k}| = |\boldsymbol{k}|$ and $d\boldsymbol{k} = d(R^{-1}\boldsymbol{k})$ since the Jacobian of a rotation is 1, then we get:

$$\mathcal{F}_{\widetilde{\boldsymbol{k}}}^{-1}(R\boldsymbol{r}) = \frac{1}{(2\pi)^3} \int d\boldsymbol{k} \frac{\boldsymbol{k}}{|\boldsymbol{k}|} \exp(-i\boldsymbol{k} \cdot R\boldsymbol{r}) = \frac{1}{(2\pi)^3} R \int d(R^{-1}\boldsymbol{k}) \frac{R^{-1}\boldsymbol{k}}{|R^{-1}\boldsymbol{k}|} \exp(-iR^{-1}\boldsymbol{k} \cdot \boldsymbol{r}) = R\mathcal{F}_{\widetilde{\boldsymbol{k}}}^{-1}(\boldsymbol{r}) \qquad \text{[S14]}$$

The value of the integral thus rotates in the same way as the position.

Next we look at the integrals behavior when scaling $\boldsymbol{r}$ with $t > 0$. Here we use that $\frac{\boldsymbol{k}}{|\boldsymbol{k}|} = \frac{t\boldsymbol{k}}{|t\boldsymbol{k}|}$ and $d(t\boldsymbol{k}) = t^3 d\boldsymbol{k}$ to get:

$$\mathcal{F}_{\widetilde{\boldsymbol{k}}}^{-1}(t\boldsymbol{r}) = \frac{1}{(2\pi)^3} \int d\boldsymbol{k} \frac{\boldsymbol{k}}{|\boldsymbol{k}|} \exp(-i\boldsymbol{k} \cdot t\boldsymbol{r}) = \frac{1}{(2\pi)^3} t^{-3} \int d(t\boldsymbol{k}) \frac{t\boldsymbol{k}}{|t\boldsymbol{k}|} \exp(-it\boldsymbol{k} \cdot \boldsymbol{r}) = t^{-3}\mathcal{F}_{\widetilde{\boldsymbol{k}}}^{-1}(\boldsymbol{r}) \qquad \text{[S15]}$$

The length of the value of the integral thus scales as $|\boldsymbol{r}|^{-3}$.

As we now know how the integral behaves under rotation and scaling of $\boldsymbol{r}$, we now only need to solve the integral at two points, one being zero, and then use the found symmetry to get the full $\boldsymbol{r}$-dependence. We choose $\boldsymbol{r}_0 = (0,0,1)$ :

$$\mathcal{F}_{\widetilde{\boldsymbol{k}}}^{-1}(\boldsymbol{r_0}) = \frac{1}{(2\pi)^3}\int d\boldsymbol{k}\frac{\boldsymbol{k}}{|\boldsymbol{k}|}\exp(-i\boldsymbol{k_3}) \quad \text{[S16]}$$

Changing to polar coordinates:

$$\mathcal{F}_{\widetilde{\boldsymbol{k}}}^{-1}(\boldsymbol{r_0}) = \frac{1}{(2\pi)^3}\int_0^\infty k^2 dk \int_0^\pi \sin\theta\, d\theta \int_0^{2\pi} d\phi \begin{pmatrix}\sin\theta\cos\phi \\ \sin\theta\sin\phi \\ \cos\theta\end{pmatrix} \cdot \exp(-i\cos\theta\cdot k) \quad \text{[S17]}$$

The $\phi$ and $\theta$ parts are easily solved, leaving us with:

$$\mathcal{F}_{\widetilde{\boldsymbol{k}}}^{-1}(\boldsymbol{r_0}) = \frac{i}{2\pi^2}\int_0^\infty dk \begin{pmatrix}0\\0\\1\end{pmatrix}\cdot(k\cdot\cos k - \sin k) \quad \text{[S18]}$$

This is solved by taking the limit:

$$\mathcal{F}_{\widetilde{\boldsymbol{k}}}^{-1}(\boldsymbol{r_0}) = \frac{i}{2\pi^2}\lim_{a\to 0}\int_0^\infty dk \begin{pmatrix}0\\0\\1\end{pmatrix}\cdot\exp(-ak)\,(k\cdot\cos k - \sin k) \quad \text{[S19]}$$

By rewriting cos and sin using exponentials and using $\int_0^\infty x^n \exp(-ax)\,dx = \frac{\Gamma(n+1)}{a^{n+1}}$, the integral is solved and the limit $a \to 0$ taken, giving:

$$\mathcal{F}_{\widetilde{\boldsymbol{k}}}^{-1}(\boldsymbol{r_0}) = \begin{pmatrix}0\\0\\-\frac{i}{\pi^2}\end{pmatrix} \quad \text{[S20]}$$

For $\boldsymbol{r} = (0,0,0)$ the $\theta$ part of the integral is easily seen to give the zero vector, $\boldsymbol{0}$.

Using the symmetry properties we can now write the general form of the integral as:

$$\mathcal{F}_{\widetilde{\boldsymbol{k}}}^{-1}(\boldsymbol{r}) = \mathcal{F}^{-1}[\widetilde{\boldsymbol{k}}] = \begin{cases}-\frac{i}{\pi^2}\frac{\boldsymbol{r}}{|r|^4}, & |r| \neq 0 \\ \boldsymbol{0}, & |r| = 0\end{cases} \quad \text{[S21]}$$

It is more convenient to move all constant outside of the function and define

$$\boldsymbol{\Upsilon}(\boldsymbol{r}) = \begin{cases}\frac{\boldsymbol{r}}{|r|^4}, & |r| \neq 0 \\ \boldsymbol{0}, & |r| = 0\end{cases} \quad \text{[S22]}$$

The expression for the 3D-mΔPDF can now be written, remembering that one term in the cross correlation is complex conjugate:

$$3D - m\Delta PDF = \frac{r_0^2}{4\mu_B^2}\langle \delta\boldsymbol{M}\,\overline{\otimes}\,\delta\boldsymbol{M} - \frac{1}{\pi^4}(\delta\boldsymbol{M} \,\overline{*}\, \boldsymbol{\Upsilon}) \otimes (\delta\boldsymbol{M} \,\overline{*}\, \boldsymbol{\Upsilon})\rangle \quad \text{[S23]}$$

### S2.2. Elastic and energy-integrated scattering

The equations above are given in the static approximation where the scattered signal is integrated in energy over the magnetic excitations. In that case we get the equal-time correlation function. In the case where only elastic scattering is used, the equations have a small modification.

The cross section, before given by equation [S1] is for elastic only scattering instead given by (Lovesey, 1984)

$$\frac{d\sigma}{d\Omega} = r_0^2 \langle \boldsymbol{Q}_\perp(-\boldsymbol{k}) \rangle \cdot \langle \boldsymbol{Q}_\perp(\boldsymbol{k}) \rangle \qquad \text{[S24]}$$

The result of this in that the equation for the 3D-mΔPDF instead becomes

$$3D - m\Delta PDF = \frac{r_0^2}{4\mu_B^2} \langle \delta\boldsymbol{M} \rangle \overline{\otimes} \langle \delta\boldsymbol{M} \rangle - \frac{1}{\pi^4} ( \langle \delta\boldsymbol{M} \rangle \ \bar{*} \ \boldsymbol{\Upsilon}) \otimes (\langle \delta\boldsymbol{M} \rangle \ \bar{*} \ \boldsymbol{\Upsilon}) \qquad \text{[S25]}$$

Which is the same as before, but for the time averaged magnetization density.

In cases where the system does not have significant magnetic excitations such as magnons, the two equations are the same, as the time-average makes no difference.

This is the case for bixbyite, where the magnetic diffuse scattering is elastic. This can be seen in Figure S4 where the total energy integrated scattering for bixbyite is shown for the data before the elastic discrimination is used for 300K and 7K. It is seen that the magnetic signal which appears at 7K is the same as was seen in figure 3, where the elastic-only contribution is shown.

The two cases, equation [S23] and [S25] are two extremes where full energy integration or purely elastic scattering is used. In practice all experiments have a finite width for the energy integration. This integration width determines which modes are time-averaged in the 3D-mΔPDF, as in equation [S25] and which give an equal-time correlation as in equation [S23]. All modes with higher energy than the integration width will be averaged in the resulting 3D-mΔPDF.

The CORELLI instruments elastic resolution changes with the scattering vector, and the resolution is better for short scattering vectors than long. The scattering vector dependency of the energy resolution is discussed by (Ye et al., 2018), where they find the energy resolution to range from 0.4 to 2.5 meV for a collected dataset.

**Table S1** Structure refinement results from single-crystal neutron diffraction data for bixbyite.

| Temperatures [K] | 7 | 25 | 50 | 80 | 160 | 240 | 300 |
|---|---|---|---|---|---|---|---|
| Unit cell length [Å] | 9.409(1) | 9.409(1) | 9.399(1) | 9.402(1) | 9.401(1) | 9.411(1) | 9.401(1) |
| 24*d* site Mn occupancy | 0.508(3) | 0.508(3) | 0.506(4) | 0.508(3) | 0.507(3) | 0.509(3) | 0.510(4) |
| 8*b* site Mn occupancy | 0.243(6) | 0.245(5) | 0.247(6) | 0.250(5) | 0.247(5) | 0.244(5) | 0.245(6) |
| 24*d* site z-coordinate | 0.4585(1) | 0.45831(1) | 0.45833(2) | 0.45831(1) | 0.45822(1) | 0.45808(1) | 0.45829(2) |
| Oxygen x-coordinate | 0.61514(7) | 0.61512(7) | 0.61517(8) | 0.61517(7) | 0.61519(8) | 0.61519(7) | 0.61531(8) |
| Oxygen y-coordinate | 0.83768(6) | 0.83771(7) | 0.83771(7) | 0.83771(7) | 0.83769(7) | 0.83767(7) | 0.83767(7) |
| Oxygen z-coordinate | 0.60997(7) | 0.60998(7) | 0.61001(8) | 0.60996(7) | 0.60994(7) | 0.60990(7) | 0.60983(7) |
| Component of first twin | 0.522(4) | 0.458(4) | 0.473(4) | 0.455(4) | 0.459(4) | 0.430(4) | 0.466(4) |
| Goodness of Fit | 1.250 | 1.195 | 1.210 | 1.183 | 1.219 | 1.220 | 1.251 |
| R1 | 0.0405 | 0.0426 | 0.0454 | 0.0441 | 0.0429 | 0.0418 | 0.0428 |
| wR | 0.1097 | 0.1094 | 0.117 | 0.113 | 0.1127 | 0.1061 | 0.1161 |

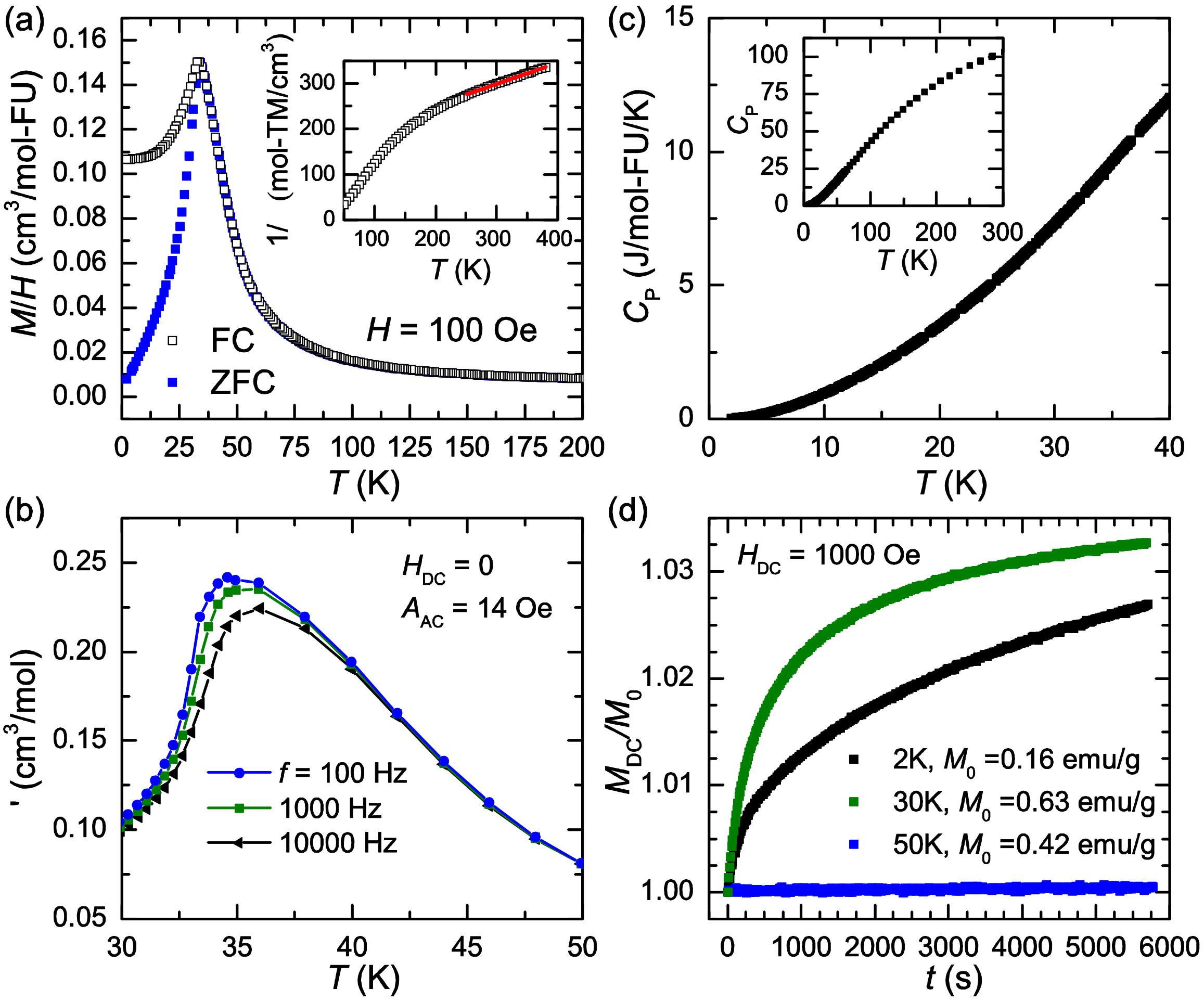

**Figure S1** Physical property measurements on bixbyite. (a) Temperature-dependent magnetization data for field-cooled (FC) and zero field cooled (ZFC) measurements showing a divergence between the two datasets. (b) in-phase component of the ac susceptibility showing frequency dependence of the broad peak near 32.5K (c) Specific heat capacity data do not reveal any clear anomaly associated with the magnetic transition near 32.5K. (d) Time-dependence of the dc magnetization data at various temperatures demonstrating glassy dynamics below the freezing temperature.

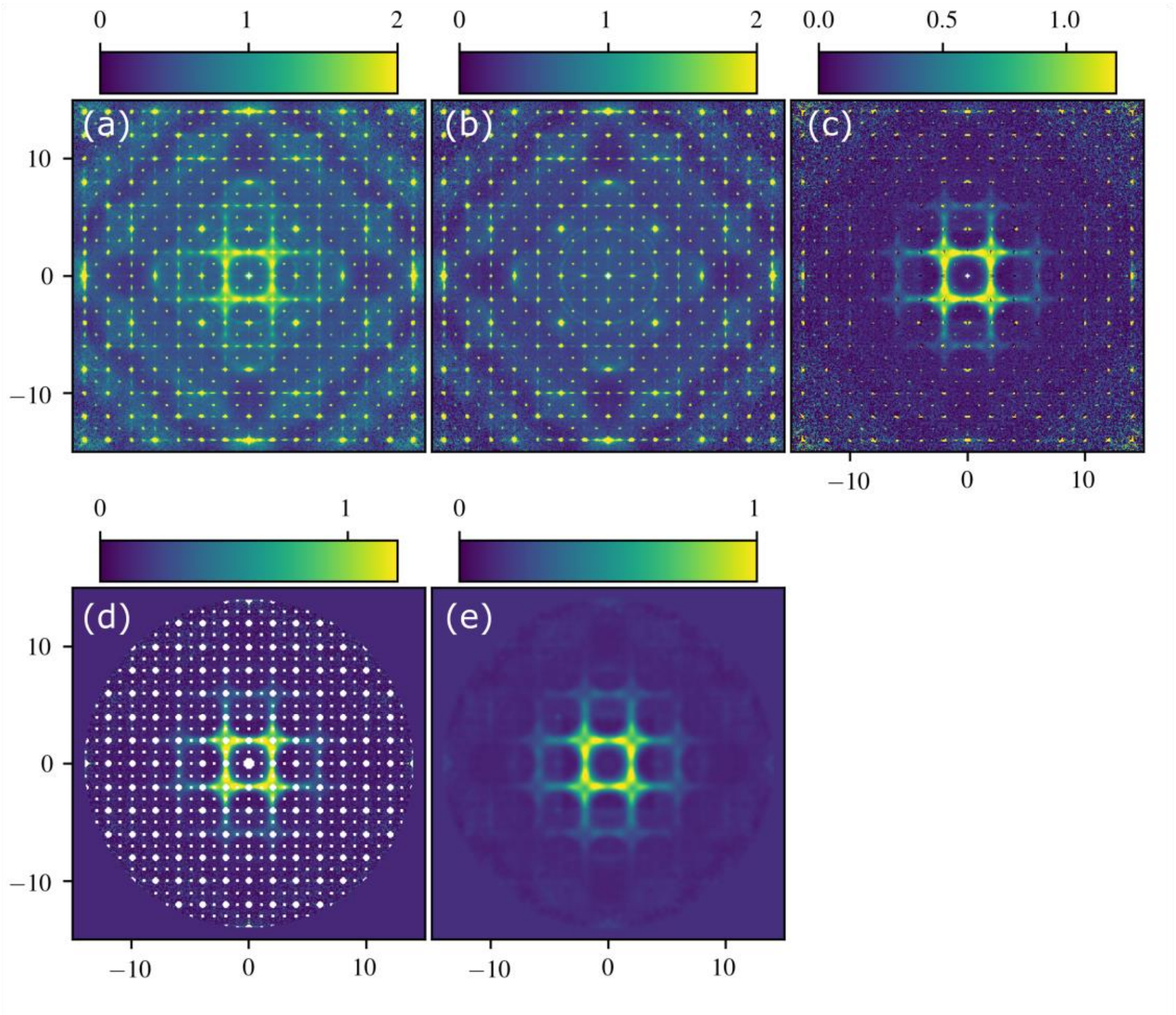

**Figure S2** Process from scattering data to isolated magnetic diffuse scattering. (a) Total elastic neutron scattering at 7K. (b) Total elastic neutron scattering at 300K. (c) Difference scattering obtained by subtracting the 300K data from the 7K data. (d) Residual intensities at Bragg positions and high angle noise removed. (e) Isolated magnetic diffuse scattering after filling in the removed Bragg areas with a smooth function resembling the diffuse scattering.

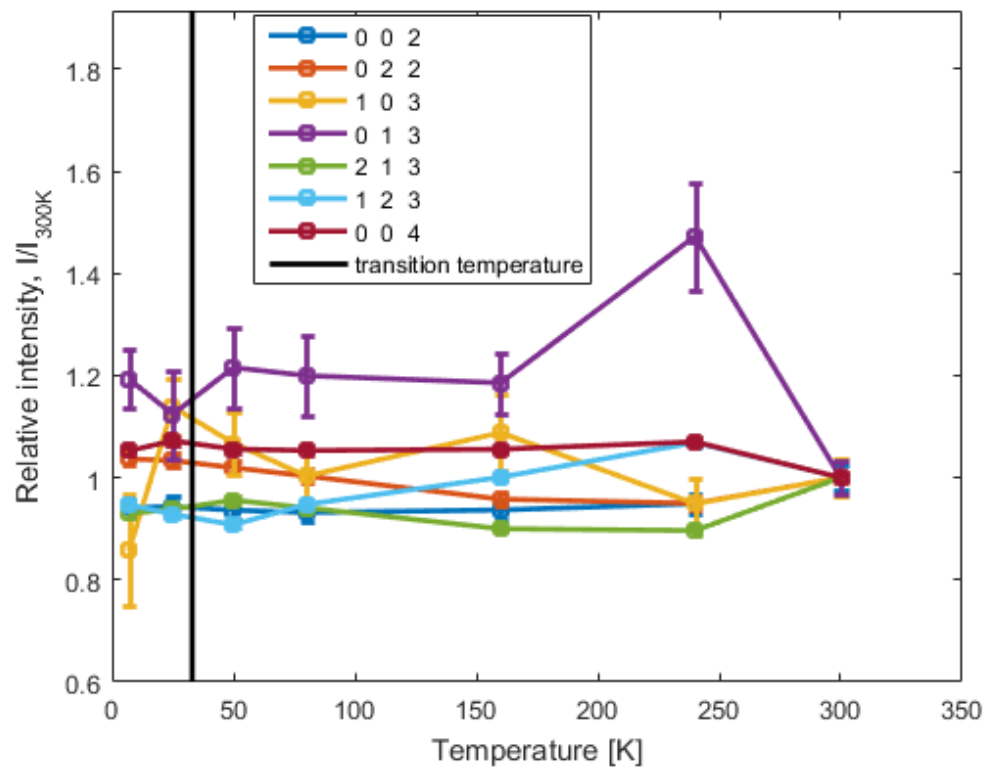

**Figure S3** **Normalized low order reflection intensities as a function of temperature**. No systematic changes are seen at the transition temperature (Vertical black line.)

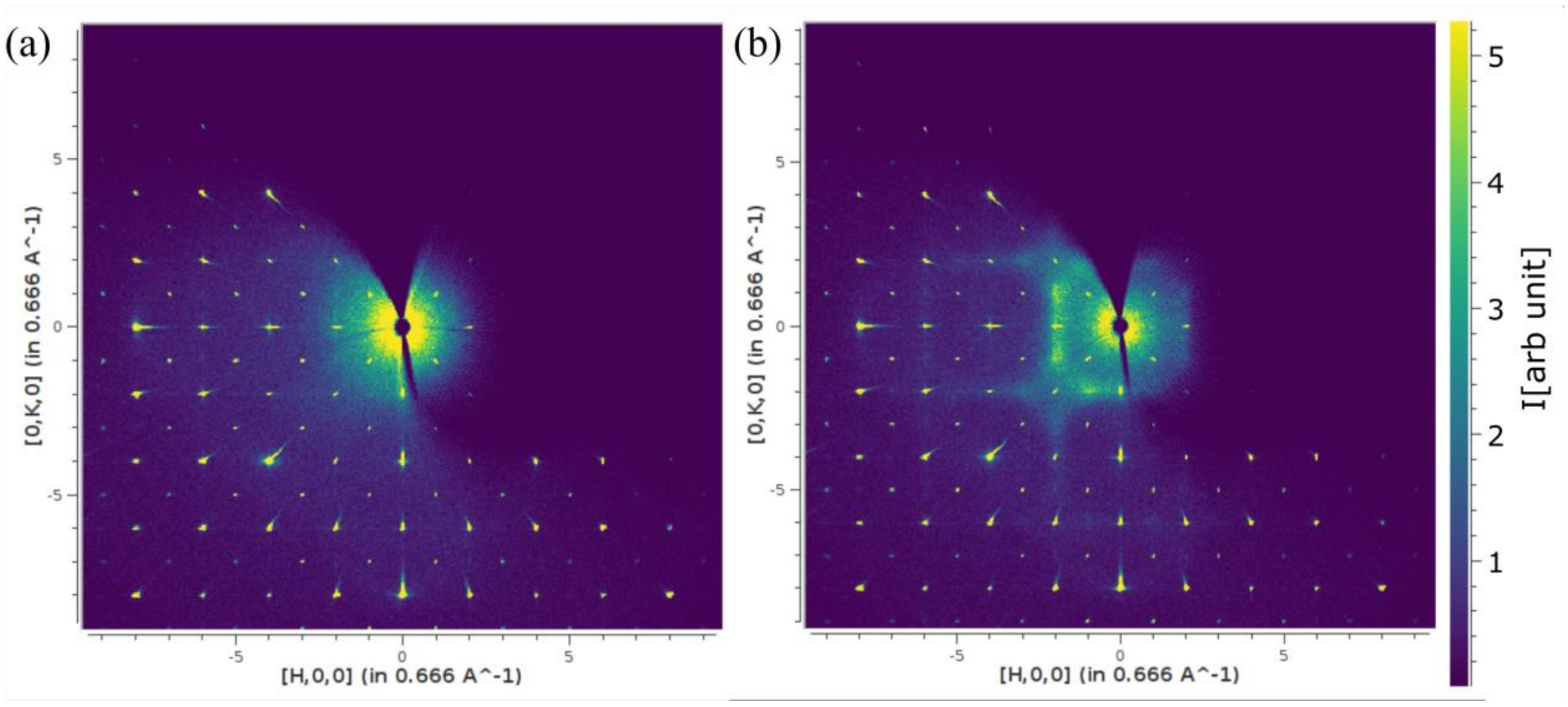

**Figure S4** Energy integrated scattering from bixbyite before corrections and elastic-only discrimination for (a) 300K and (b) 7k. The magnetic diffuse scattering appearing at 7K is the same as was seen in figure 3 where the elastic component is shown.

**S3. Additional References**